\documentstyle[12pt,iopfts]{ioplppt}
\begin{document}
\jl{3}
\title{Spin and energy correlations in the one dimensional spin 1/2 
Heisenberg model}
\author{F Naef and X Zotos}
\address{
Institut Romand de Recherche Num\'erique en Physique des
Mat\'eriaux (IRRMA), \\
EPFL-PPH, CH-1015 Lausanne, Switzerland}
\begin{abstract}
In this paper, we study the spin and energy dynamic correlations of the
one dimensional spin 1/2 Heisenberg model, using mostly exact diagonalization 
numerical techniques.  In particular, observing that
the uniform spin and energy currents decay to finite values 
at long times, we argue for the absence of spin and energy diffusion 
in the easy plane anisotropic Heisenberg model.
\end{abstract}
\maketitle

\section{Introduction}
Recently there has been a renewed interest in the finite temperature dynamics 
of the one dimensional spin 1/2 Heisenberg model, especially on the question
of diffusive spin transport\cite{bl,fab,sand,msn}. 
In particular, it was argued 
that the integrability of the model implies pathological spin dynamics and 
presumably the absence of spin diffusion\cite{zp,mcc2}. 
The role of conservation laws was pointed out in reference\cite{znp} were 
it was shown that in several quantum integrable models 
the uniform ($q=0$) current correlations do not decay to zero at long times. This 
result, established using the Mazur inequality\cite{maz}, suggests    
pathological finite temperature dynamics.

As far as the Heisenberg model is concerned, the analysis of conservation
laws has shown that the energy current operator commutes with the Hamiltonian,
suggesting anomalous finite-($q,\omega$) energy density correlations.
However, for zero magnetic field, this method turned out to be 
insufficient for deciding about the decay of the uniform spin current 
correlations.
This case is closely related to the behavior of
the finite temperature conductivity in the one dimensional model of 
spinless fermions at half-filling interacting with a nearest neighbor 
interaction (the ``t-V" model\cite{znp}).

In this work, we address the issues raised above by the numerical 
diagonalization of the Hamiltonian matrix on finite size lattices.
More precisely, we study the 
implications of the energy current 
conservation on the ($q,\omega$) energy density correlations, and,
as an alternative route to the analysis of spin diffusion, we investigate
the decay of the uniform ($q=0$) spin current correlations.

The paper is organized as follows: in section 2, we recall the  Heisenberg Hamiltonian
and define the various quantities studied below. 
In section 3 we briefly summarize the phenomenological picture of diffusion. 
There, we also argue that the decay
of the uniform spin current correlations to a finite value
is incompatible with a diffusive behavior, {\it assuming}
continuity in the wave-vector $q$ of the correlations at $q=0$. 
Next,
we test these ideas in section 4 in the XY limit, where results 
can be obtained analytically. 
Turning to the numerical results, in section 5.1
we present the 
energy density correlations at infinite temperature for the case of the 
isotropic Heisenberg model. A simple ansatz for the observed behavior 
suggests a logarithmic dependence at low frequencies for the 
energy autocorrelation function.
As far as the spin dynamics is concerned, numerous studies of the 
$(q,\omega)$ spin density correlations exist\cite{bl,fab,sand}.
Therefore, in section 5.2, we restrict ourselves to the decay of  
the uniform spin current correlations for various values of the anisotropy 
parameter $\Delta$ and temperatures. Interestingly, it turns out that these do 
not decay to zero for $\Delta<1$. According to the argument given in section 3,
this result implies non-diffusive spin transport.
Section 6 contains a short discussion on 
experimental relevance of these findings and open questions.

\section{The model}
The anisotropic Heisenberg Hamiltonian for a chain of $L$ sites 
with periodic boundary conditions is given by:
\begin{equation}
H=\sum_{l=1}^L h_l=J \sum_{l=1}^L (S_l^x S_{l+1}^x +
 S_l^y S_{l+1}^y + \Delta S_l^z S_{l+1}^z) ,
\label{heis}
\end{equation}
where $S_l^{\alpha}=\frac{1}{2}\sigma_l^{\alpha}$,
$\sigma_l^{\alpha}$ are the Pauli spin operators with components $\alpha=x,y,z$
at site $l$.

For a conserved quantity $A=\sum_{l=1}^L a_l$, $[A,H]=0$, the continuity 
equation in $q-$space defines the current $j_q$:

\begin{equation}
\frac{\partial a_q(t)}{\partial t}= 2\i\sin(q/2) j_q
\label{ce}
\end{equation}
with
\begin{equation}
a_q=\frac{1}{\sqrt L} \sum_{l=1}^L \e^{\i ql} a_l,~~~ 
j_q=\frac{1}{\sqrt L} \sum_{l=1}^L \e^{\i ql} j_l
\end{equation}
and $a_q(t)=\e^{\i Ht} a_q \e^{-\i Ht}$.

Setting $a_l=S^z_l, h_l$ we find the following spin and energy currents 
respectively:
\begin{equation}
j^z_l=J (S_l^y S_{l+1}^x-S_l^x S_{l+1}^y)
\label{jz}
\end{equation}
\begin{eqnarray}
j^H_l & = & J^2 (S_{l-1}^x S_l^z S_{l+1}^y-S_{l-1}^y S_l^z S_{l+1}^x)  \nonumber\\
 & + & J^2\Delta (S_{l-1}^y S_l^x S_{l+1}^z-S_{l-1}^z S_l^x S_{l+1}^y) \nonumber\\
 & + & J^2\Delta (S_{l-1}^z S_l^y S_{l+1}^x-S_{l-1}^x S_l^y S_{l+1}^z)
\label{je}
\end{eqnarray}
For the discussion of dynamic correlations at finite temperatures, we 
chose to analyze the anticommutator form:
\begin{equation}
S_{AA}(q,t-t')=\frac{1}{2} \langle\{a_q(t),a_{-q}(t')\}\rangle
\label{sqt}
\end{equation}
where $\langle\;\rangle$ is the thermal average at temperature
$T=1/\beta$ over a complete set of states.  

Further, the frequency dependent correlation function defined by:
\begin{equation}
S_{AA}(q,\omega)=\int_{-\infty}^{+\infty} \d\omega \e^{\i\omega t} S_{AA}(q,t)
\label{sqo}
\end{equation}
is symmetric in frequency, $S_{AA}(q,\omega)=S_{AA}(q,-\omega)$.

A central point in our approach is the relation between the dynamic 
correlations
of a quantity A and its corresponding current correlations, which we obtain by 
using the continuity equation (\ref{ce}):
\begin{equation}
\omega^2 S_{AA}(q,\omega)=4\sin^2(q/2) S_{j^Aj^A}(q,\omega)
\label{cs}
\end{equation}

In particular, we will discuss the asymptotic value of the current 
correlations
\begin{equation}
C_{j^Aj^A}=\lim_{t\rightarrow \infty} 
\frac{S_{j^Aj^A}(q=0,t)}{S_{j^Aj^A}(q=0,t=0)}. 
\end{equation}

A finite value of $C_{j^Aj^A}$ translates to a $\delta(\omega)$ peak in 
$S_{j^Aj^A}(q=0,\omega)$ and, as we will discuss below, implies restrictions 
in the behavior of $S_{AA}(q,\omega)$. 

An important observation is that the energy 
current $j^H$ of the Heisenberg model commutes
with the Hamiltonian\cite{znp}, 
so that $C_{j^Hj^H}=1$, whereas the spin current does not. However, 
it will turn out that $C_{j^zj^z} > 0$ for $\Delta<1$, meaning that
the spin current and energy current correlations are
similar in the sense that in their frequency representation, they both exhibit
a finite weight $\delta(\omega)$ function.

\section{Diffusive behavior} 
When we consider the $(q,\omega)$-dependent correlations of a 
conserved quantity $A$
such as the magnetization, it is usually assumed, largely on phenomenological 
grounds, that they exhibit a diffusive behavior in the long-time
$|t-t'|\rightarrow\infty$, short wavelength $q\rightarrow 0$ regime\cite{km}: 
\begin{equation}
S_{AA}(q,t-t')\sim \e^{-D_Aq^2|t-t'|}
\label{db}
\end{equation}
where $D_A$ is the corresponding diffusion constant, or     
\begin{equation}
S_{AA}(q,\omega)\sim\frac{2D_Aq^2}{(D_Aq^2)^2+\omega^2}
\label{dbqo}
\end{equation}
for $\omega\rightarrow 0$.

This Lorentzian form correctly reduces to a $\delta(\omega)$ function  
in the limit $q\rightarrow 0$, as implied by $[A,H]=0$. Further, 
using the continuity 
equation (\ref{cs}) for $q\rightarrow 0$, we obtain:
\begin{equation}
S_{j^Aj^A}(q,\omega)\sim\frac{2D_A\omega^2}{(D_Aq^2)^2+\omega^2}
\label{djbqo}
\end{equation}
which gives the diffusion constant $D_A$ when first, the limit  
$q\rightarrow 0$ and then, $\omega \rightarrow 0$ are taken.
On the other hand, if the current correlations for $q=0$
do not decay to zero at 
long times, $C_{j^Aj^A}> 0$ and $S_{j^Aj^A}(q,\omega)$ has a finite 
weight $\delta(\omega)$ component
which is incompatible with the diffusive form 
(\ref{djbqo}). In this reasoning, we must assume a regular behavior  
of the correlation functions in the $q$ variable.

To summarize the argument, if a quantity $A$ is conserved ($[A,H]=0$) and 
its current $j^A$ is either conserved ($[j^A,H]=0$), or $C_{j^Aj^A}> 0$, 
then continuity in $q$ at $q=0$ excludes a diffusive form (\ref{db}) for the 
corresponding correlation $S_{AA}(q,t-t')$. 

\section{XY limit} 
A simple model for testing these ideas is the $XY$ limit 
($\Delta=0$), of the Heisenberg model. In this case, both the 
energy current $j^H$ and the spin current $j^z$ commute with the 
Hamiltonian. The model can be mapped to a free spinless fermion model by
using a Jordan-Wigner transformation which allows us also to evaluate 
explicitly the spin and energy dynamic correlations at $\beta=0$. In the spin
case, these are well known results\cite{nie}:
\begin{equation}
S_{S^z S^z}(q,\omega)=\frac{1}{2\,(4\,J^2\,\sin^2(\frac{q}{2})-\omega^2)^{1/2}}
\; \theta(|2\,J\,\sin(\frac{q}{2})|-|\omega|)
\label{xysz}
\end{equation}
\begin{equation}
S_{HH}(q,\omega)=\frac{(4\,J^2\,\sin^2(\frac{q}{2})-\omega^2)^{1/2}}
{8\,\sin^2(\frac{q}{2})}\;
\theta(|2\,J\,\sin(\frac{q}{2})|-|\omega|)
\label{xye}
\end{equation}
These forms are indeed consistent with the conservation of
both spin (energy) and spin current (energy current) as they reduce
to a $\delta(\omega)$ function when the limits $q\rightarrow 0$, 
$\omega \rightarrow 0$ are taken.

Further, the time decay of the autocorrelations at $\beta=0$ is not of 
the form $1/\sqrt{t}$,
as predicted by the diffusion hypothesis. Indeed,
\begin{equation}
\langle S^z_l(t) S^z_l \rangle = \frac{1}{4} \,J_0^2(Jt)
\label{xyszt}
\end{equation}
\begin{equation}
\langle h_l(t) h_l \rangle = \frac{J^2}{8} \,\left( J_0^2(Jt)+J_1^2(Jt) \right)
\label{xyet}
\end{equation}
which both behave as $1/t$ for $t\rightarrow\infty$.

\section{Anisotropic Heisenberg model}
\subsection{Energy correlations}
As we mentioned earlier, the energy current $j^H$ associated with
the anisotropic Heisenberg model (\ref{heis}) 
commutes with the Hamiltonian for all values of 
the parameter $\Delta$. Therefore, the time correlations do not 
decay at all ($C_{j^Hj^H}=1$) and according to the argument explained 
in section 3, no 
diffusive energy transport occurs. However, the conservation of $j^H$ does not
provide us with any details about the shape of 
$S_{HH}(q,\omega)$ at finite $q$. 
In the absence of an analytical solution, we investigate this quantity by
numerical diagonalization of the Hamiltonian matrix on a ring of 16 sites.

In figure 1, we show $S_{HH}(q,\omega)$ for $\Delta=1$, which is experimenally
the most interesting point as it describes isotropic quasi one-dimensional 
antiferromagnets. We study the high temperature limit $\beta=0$, which is 
the most convenient for a numerical study as it involves the full excitation 
spectrum, but is also relevant experimentally for spin systems as the
magnitude of $J$ can be of the order of $T$.
The plot is represented as histograms of width $0.06 \omega/J$, all the 
frequencies which fall into one interval are summed up. The inset shows the 
normalized, integrated (prior to summing nearby frequencies) quantity 
\begin{equation}
I_{HH}(q,\omega)=\frac{\int_{0+}^{\omega} \d\omega' S_{HH}(q,\omega')}
{\int_{0+}^{\infty} \d\omega' S_{HH}(q,\omega')}
\end{equation}
which has the advantage of smoothing out the finite size discontinuities. 
To point out the practically linear integrated behavior of the
pure Heisenberg model, we also show the same quantity for a more generic case
obtained by adding a next-next neighbor (non-integrable) interaction $J_2$.

The simplest way to describe this behavior is by means of ``plateaus" 
given by the following ansatz:
\begin{equation}
S_{HH}(q,\omega)=\frac{\sqrt{3}\pi J}{16\sqrt{1-\cos(q)}}\,
\theta(|\omega|-J\sqrt{3(1-\cos(q))}
\end{equation}
which satisfy the first $\int \d\omega\,S_{HH}(q,\omega)=3\pi J^2/8$ and the 
second $\int \d\omega\,\omega^2S_{HH}(q,\omega)=3\pi J^4(1-\cos(q))/8$ exact 
moments for $\beta=0$. Further, this ansatz is compatible with the limit 
$S_{HH}(q\rightarrow 0,\omega)\rightarrow \delta(\omega)$ as implied
by the conservation of energy. 
Using the continuity equation (\ref{cs}), we obtain for small $q$ and $\omega$
\begin{equation}
S_{j^Hj^H}(q,\omega)=\frac{\sqrt{6}\pi J}{16}\,
\frac{\omega^2}{q^3}\,\theta(|\omega|-\sqrt{\frac{3}{2}J}|q|) 
\end{equation}
which correctly reduces to a $\delta(\omega)$-function for $q\rightarrow 0$, 
in agreement with the conservation of the energy current $[j^H,H]=0$.

Using this ansatz we find for the energy autocorrelation function 
(obtained by integration over $q$):
\begin{equation}
\int_{-\infty}^{+\infty}\langle h_l(t) h_l \rangle \,\e^{\i\omega t}\,
d t=C_0-C_1\ln(\omega/J)+O(\omega^2)\;,\hspace{0.5cm}
C_0,C_1>0
\label{eauto}
\end{equation}
a logarithmic behavior at low frequencies, 
in contrast to the diffusion form $\frac{1}{\sqrt{\omega}}$.
 
We should stress that these results are only {\it indicative}, as they are  
obtained from small size lattices which can provide reliable information 
only for correspondingly high frequencies and large wave-vectors. 
Nevertheless, the consistency of these results with the arguments presented 
above against a diffusion form are encouraging.

\subsection{Spin correlations} 
The spin density dynamic correlations $S_{S^zS^z}(q,\omega)$ have been 
the subject of many studies which have not been able to
answer the question of spin diffusion unambiguously. 
Here, we revisit this problem by investigating
the compatibility between spin density and spin current correlations, 
which requires that we calculate $C_{j^zj^z}$.
In contrast to the energy current, the spin current $j^z$ does not
commute with the Hamiltonian, so that $S_{j^zj^z}(q=0,\omega)$ is different 
from a pure $\delta(\omega)$ function. Nevertheless, if $C_{j^zj^z}>0$, 
which means that $S_{j^zj^z}(q=0,\omega)$ has a finite weight 
$\delta-$function at $\omega=0$,
our previous arguments against diffusion still hold.

In determining $C_{j^zj^z}$, 
we noticed a peculiar difference in the low 
frequency behavior of $S_{j^zj^z}(q=0,\omega)$ 
depending on the anisotropy parameter $\Delta$. In figure 2, 
we show
\begin{equation}
I_{j^zj^z}(\omega)=C_{j^zj^z}+2\int_{0+}^{\omega} \d\omega'
S_{j^zj^z}(q=0,\omega'), 
\end{equation}
the corresponding integrated, normalized quantity. 
We see that for $\Delta=\cos(\pi/n)$, 
$n=3,4,\ldots$ ($n=3$ in the figure)
all the low frequency weight of $S_{j^zj^z}(q=0,\omega)$ is concentrated 
in the $\delta$-function at $\omega=0$. 
In contrast, for neighboring values
such as $\Delta=0.45$ or $0.55$, we observe a shift of weight to
a low frequency region whose size decreases as the system grows (inset) and
eventually vanishes as $L\rightarrow \infty$.
We believe that the behavior of this
special $\Delta$ points is related to the existence of finite length
strings (bound states) as they  appear in the formulation of 
the thermodynamics of the Heisenberg
model, within the Bethe ansatz method\cite{tak}.
It seems that in order
to determine $C_{j^zj^z}$ from finite size systems for 
$\Delta\neq \cos(\pi/n)$, we should include the weight from these low
frequency regions. As an example, doing so for $\Delta=0.45$
gives us a value of $C_{j^zj^z}=0.66$ for L=16 (figure 2).
Having discussed this technical issue, we can then determine $C_{j^zj^z}$
for different size systems, as a function of temperature and $\Delta$.
By extrapolating our finite size results to the thermodynamic limit 
using second order polynomials in $1/L$ for $L=8,\cdots,18$, we obtain the 
results shown in figure 3. 
Their striking feature is that
for $T>J$, $C_{j^zj^z}$ is finite in
the $\Delta<1$ region, and practically zero when $\Delta \ge 1$. 
In this regime, according to our previous argument, we expect a non-
diffusive behavior. 

Deciding about the behavior of $C_{j^zj^z}$ for $\Delta \ge 1$ at finite 
temperatures is rather subtle.
The reason is that in the 
Heisenberg model, $\Delta=1$ corresponds to a point of change of symmetry, 
from easy plane to easy axis, accompanied by the opening of a gap. 
In the fermionic version of the model, the ``t-V" model, it corresponds to a 
metal-insulator Mott-Hubbard type transition, with the charge stiffness 
changing discontinuously\cite{ss} at zero temperature. We should note that 
this discontinuity is difficult to reproduce 
by numerical simulations on small finite size lattices, as the 
transition corresponds to the divergence of the localization length. 
Considering that at high temperature, $C_{j^zj^z}$ behaves 
similarly to the charge stiffness
in the ``t-V" model\cite{znp} we understand why 
it is difficult to decide whether $C_{j^zj^z}$ 
is greater than zero in the region $\Delta \gtrsim 1$ 
and $T<\infty$. For the same reason, we cannot exclude that $C_{j^zj^z}$ 
behaves discontinuously at $\Delta=1$. Nevertheless, 
it seems
unambiguous that $C_{j^zj^z}\simeq 0$ for $\Delta>1.5$ and $T>J$.

\section{Discussion}
The results presented are of interest in recent experimental 
studies\cite{tag} of spin dynamics in quasi-one dimensional materials 
such as $CuGeO_3$ and $Sr_2CuO_3$. Particular attention should be paid to 
the unusually high value of the diffusion constant found in NMR experiments 
on $Sr_2CuO_3$\cite{tag}, perhaps related to the integrability of the 
Heisenberg model as discussed above.
Furthermore, our results on the behavior of energy density correlations  
are of interest in the interpretation of the quasi-elastic 
Raman scattering, related to magnetic energy 
fluctuations\cite{kur,hal}. We should emphasize that no diffusion form
should be expected for the energy density correlations in the
isotropic Heisenberg model with only nearest neighbor interaction.
An eventual diffusive behavior should be attributed to next-nearest
neighbor coupling, interaction with phonons or deviations from one
dimensionality. Finally,
the main unresolved issue in this work is a better understanding 
of the finite temperature spin dynamics at the isotropic 
point. 

\ack
We would like to thank P. Prelov\v sek for useful discussions. This 
work was supported by the Swiss National Science Foundation grant 
No. 20-49486.96, the University of Fribourg and the University of Neuch\^atel.

\Figures
\Figure{ Energy density correlation function $S_{HH}(q,\omega)$ at 
$\beta =0$ for $\Delta=1$, $q=(2\pi/16) n$, $n=1,\ldots,4$.
The inset shows the normalized, integrated quantity $I_{HH}(q,\omega)$ 
for $\Delta=1$, $J_2=0$ and $J_2=0.2J$.}

\Figure{Integrated $q=0$ energy current correlations $I_{j^zj^z}
(\omega)$ for $N=16$, $\beta=0$ and $\Delta=0.45,0.5$ and $0.55$. The inset
displays $I_{j^zj^z}(\omega)$ for $L=12,14,16,18$ at inverse 
temperature $\beta=0$ 
and $\Delta=0.2$.}

\Figure{Values of $C_{j^zj^z}$ as function of $\Delta$ and $\beta$ obtained 
by extrapolating second order polynomials in $1/L$ from results on systems
of sizes $L=8,\ldots,18$. Calculations were done for $\Delta=\cos(\pi/n)$,
$n=3,4,5,6,7,10$ and $\Delta=1.0,1.1,1.5$.}

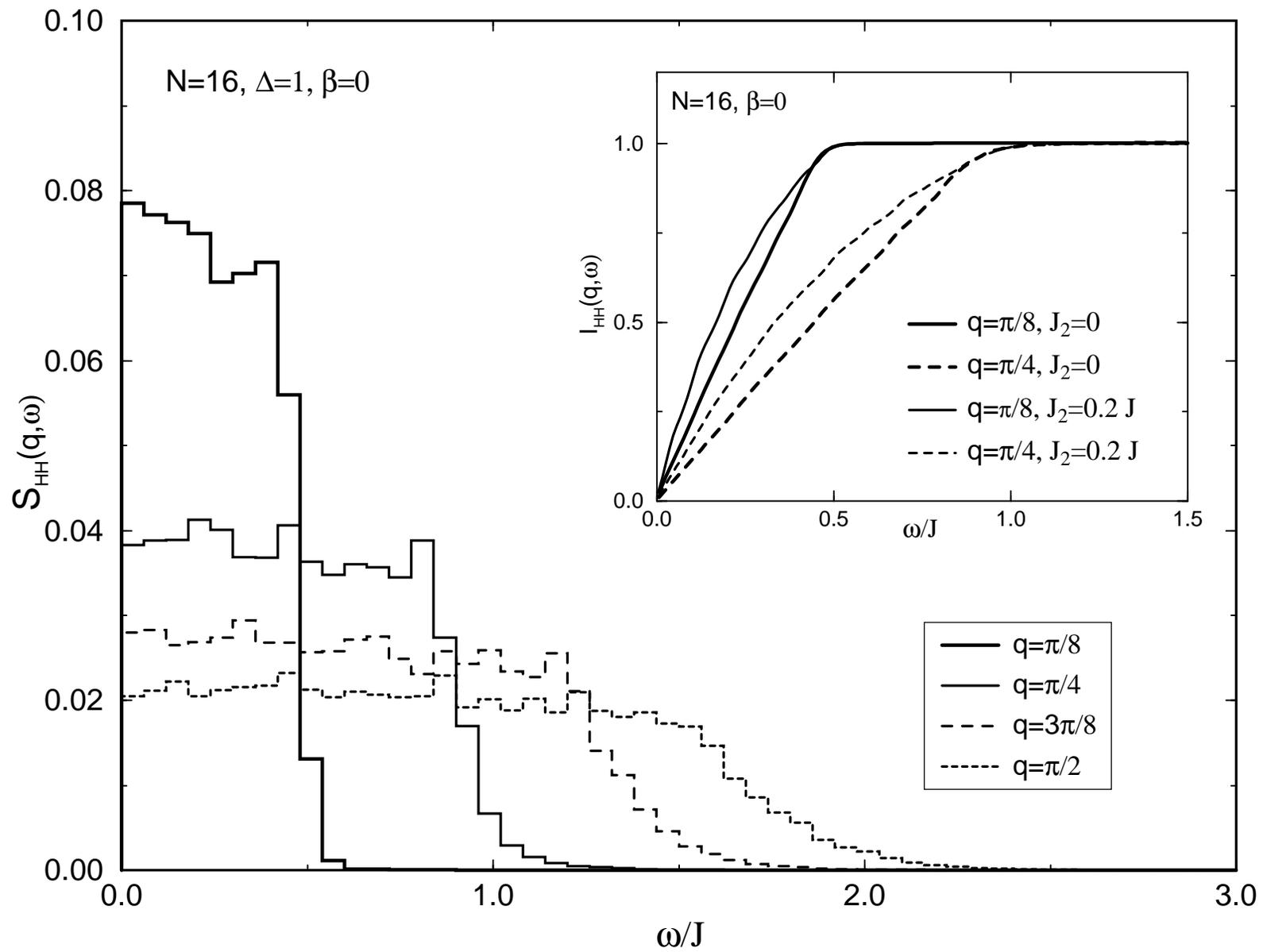
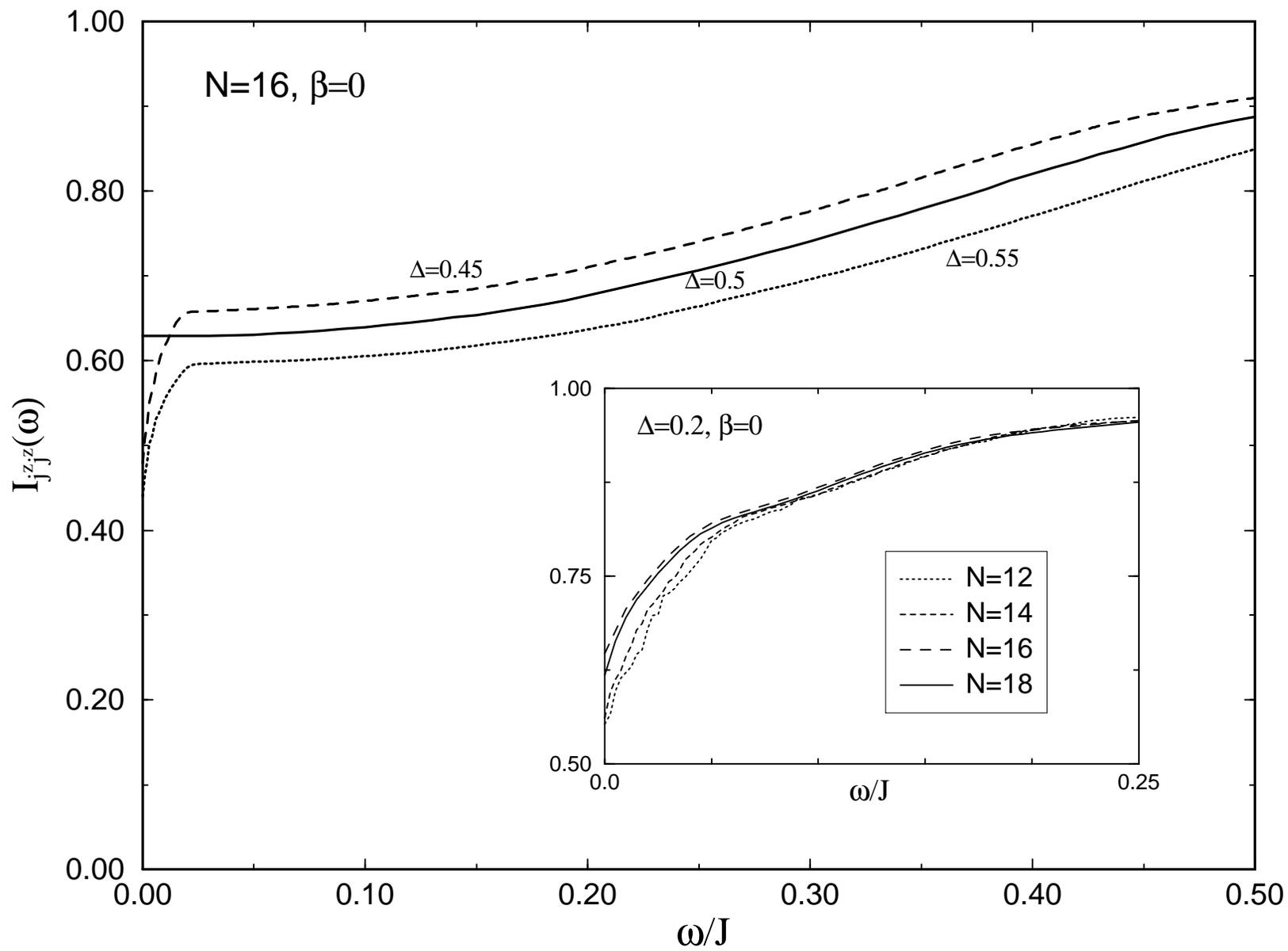
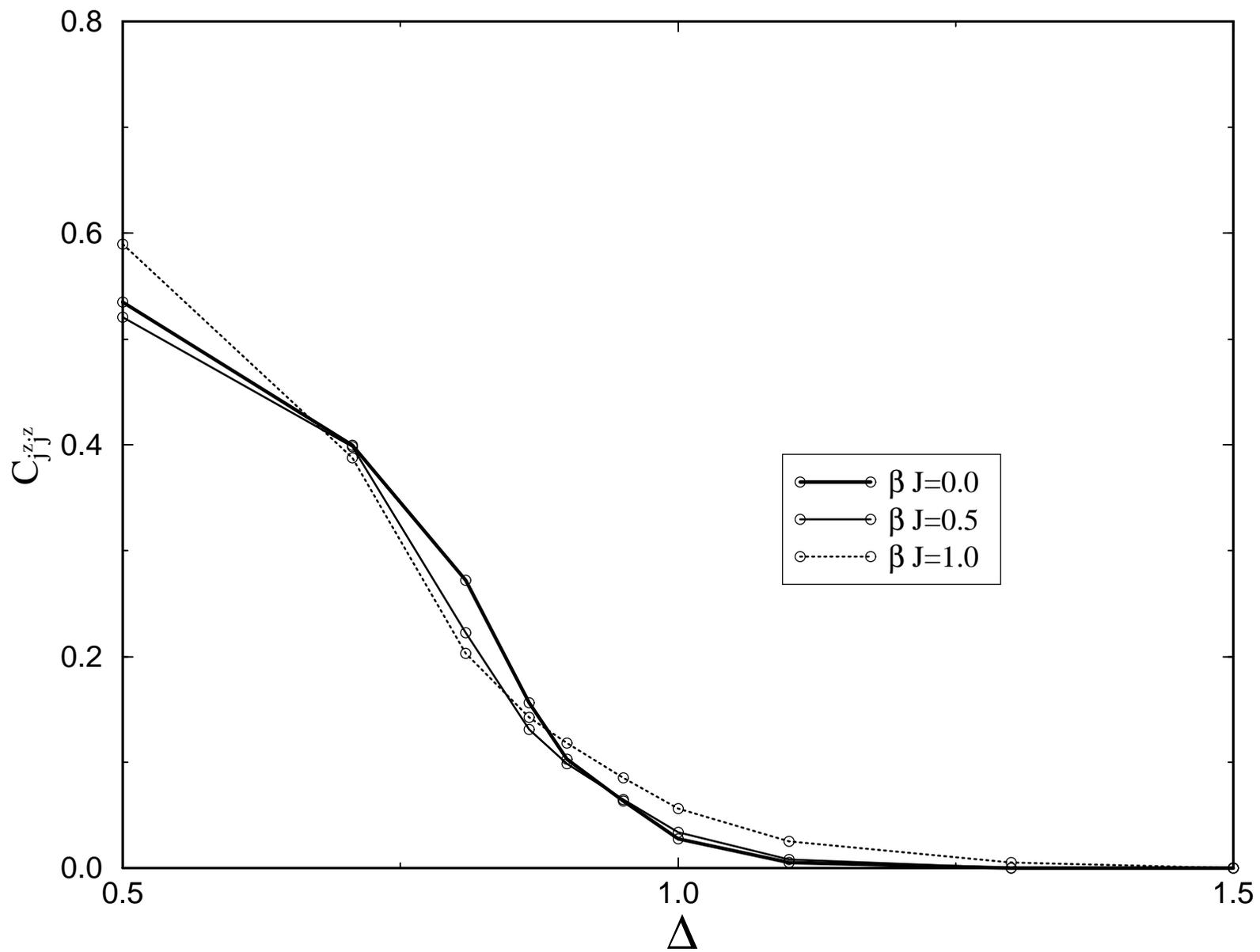
\end{document}